# Topological state engineering via supersymmetric transformations

Gerard Queraltó[1†], Mark Kremer[2†], Lukas J. Maczewsky[2], Matthias Heinrich[2], Jordi Mompart[1], Verònica Ahufinger[1], and Alexander Szameit[2*]

[1]*Departament de Física, Universitat Autònoma de Barcelona, E-08193 Bellaterra, Spain*
[2]*Institut für Physik, Universität Rostock, Albert-Einstein-Straße 23, 18059 Rostock, Germany*
[†]These authors contributed equally to this work.
*alexander.szameit@uni-rostock.de

The quest to explore new techniques for the manipulation of topological states simultaneously promotes a deeper understanding of topological physics, and is essential in identifying new ways to harness their unique features. Here, we examine the potential of supersymmetric (SUSY) transformations to systematically address, alter and reconfigure the topological properties of a system. To this end, we theoretically and experimentally study the changes that topologically protected states in photonic lattices undergo as SUSY transformations are applied to their host system. In particular, we show how SUSY-induced phase transitions can selectively suspend and re-establish topological protection of specific states. Furthermore, we reveal how understanding the interplay between internal symmetries and the symmetry constraints of supersymmetric transformations provides a roadmap to directly access the desirable topological properties of a system. Our findings pave the way for establishing SUSY-inspired techniques as a powerful and versatile tool for topological state engineering.

Physical laws are intrinsically connected to symmetries, which can be classified in spacetime and internal symmetries. Unlike any other symmetry, Supersymmetry (SUSY), originally developed as an extension of the Poincaré Group [1], offers a loophole to the Coleman-Mandula theorem [2], allowing the interplay of spacetime and internal symmetries in a non-trivial way [3]. Despite the lack of direct experimental evidence of SUSY in High-Energy Physics, where SUSY establishes a relation between bosons and fermions [1], some of its fundamental concepts have been successfully adapted to numerous fields such as Condensed Matter [4], Statistical Mechanics [5], non-relativistic Quantum Mechanics [6], Optics [7,8], and Cosmology [9]. In particular, SUSY provides an effective theory to describe quantum phase transitions occurring at the boundary of topological superconductors [10], where topological states characterized by topological invariants emerge [11,12]. In this work, SUSY transformations are applied to manipulate topological properties deeply connected to internal symmetries of the systems. Specifically, a new method for topological state engineering, e.g. to selectively suspend and re-establish the topological protection of a targeted state, is presented. Furthermore, it is shown how closely this behavior is linked to symmetry constraints of SUSY transformations [13], enabling these symmetries to be fully or partially preserved, or cancelled in their entirety. As SUSY transformations are tailored to their specific purpose, they imprint their characteristic signature on the topological invariants, as well as the related topological protection.

To explore the fruitful interplay between SUSY and Topology, we employ femtosecond laser written photonic lattices [14]. In recent years, the field of photonics has shed light on a plethora of phenomena stemming from topological phases (See [15, 16] and references therein), and photonic lattices have been established as a versatile experimental platform [17-20]. In a similar vein, SUSY notions have been introduced to photonics [8] to tackle the long-standing challenge of systematically shaping the modal content of highly multi-moded structures [21-28], controlling scattering characteristics [29-31], designing laser arrays [32,33], creating band gaps in extremely disordered potentials [34] and robust mid-gap states [35]. To elucidate how SUSY enables the manipulation of topological properties, we apply discrete SUSY transformations to photonic lattices embodying the simplest system with non-trivial topological properties, the Su-Schrieffer-Heeger (SSH) model [36]. Along these lines, we show that SUSY allows for the systematic breaking and recovery of symmetries of the system and thereby constitutes a powerful tool to tailor topological transitions and to manipulate the topological properties of a system.

**Results**

**Theory**

In its general quantum-mechanical formulation, unbroken SUSY connects two superpartner Hamiltonians $\mathcal{H}^{(1)}$ and $\mathcal{H}^{(2)}$, sharing a common set of eigenvalues except for the eigenvalue of the ground state of $\mathcal{H}^{(1)}$, which is removed from the spectrum of $\mathcal{H}^{(2)}$. A step forward towards a more general Hamiltonian spectrum manipulation, allowing the removal of different eigenvalues, can be achieved by applying SUSY-like discrete transformations [8]. Considering a one-dimensional lattice composed of $N$ evanescently-coupled single-mode waveguides, the system is characterized by a discrete Hamiltonian $\mathcal{H}$ given by an $N \times N$ tridiagonal matrix, with the propagation constants occupying the diagonal elements and the coupling strengths the off-diagonal elements. For the waveguide lattices here employed, light propagation along the $z$-direction can be described using coupled-mode equations [37]:

$$-\mathrm{i}\frac{d}{dz}\Psi = \mathcal{H}\Psi, \tag{1}$$

where $\Psi = (\psi_1, \ldots, \psi_N)^\mathrm{T}$, with $\psi_j$ describing the modal field amplitude in waveguide $j$. From the eigenvalue equation $\mathcal{H}\Psi_s = \lambda_s \Psi_s$ that relates the eigenfunction $\Psi_s$ and eigenvalues $\lambda_s$ of the state $s$, superpartner Hamiltonians can be obtained using the $QR$ factorization:

$$\mathcal{H}_m^{(1)} = \mathcal{H} - \lambda_m I = QR, \qquad \mathcal{H}_m^{(2)} = RQ, \tag{2}$$

where $Q$ is an orthogonal matrix ($Q^\mathrm{T} Q = I$), $R$ an upper triangular matrix, and $I$ the identity matrix [38]. The superpartner Hamiltonian $\mathcal{H}_m^{(2)}$, shares a common set of eigenvalues with $\mathcal{H}_m^{(1)}$, except for $\lambda_m$ that has been removed from the spectrum (see the right part of Fig. 1a). Note that the standard SUSY transformation annihilating the fundamental state can still be carried out with this method, as it is displayed in the left part of Fig. 1a. The corresponding eigenvalue $\lambda_m$ is removed because its eigenstate $\Psi_m$ is completely localized in the fully decoupled $N^{th}$ waveguide and, as such, does not have any influence on the dynamics of the remaining system of $N$-$1$ waveguides (See Supplementary S1 for more details). By applying these transformations in an iterative way, superpartner structures with desired eigenvalue spectra can be engineered by removing the desired number of eigenvalues, and reducing the overall system size. A question that naturally arises, yet to this date remains unexplored, is the impact of targeting a state with non-trivial topological properties. Does its removal irrevocably change the topological properties of the system?

The SSH model, one of the most prominent systems for illustrating topological physics, can be implemented using a one-dimensional lattice of evanescently coupled waveguides with two alternating couplings $c_1$ and $c_2$ ($c_1<c_2$). Whereas an infinite lattice is invariant under the exchange of couplings, the presence of edges in a finite SSH chain introduces two distinct types of edge terminations that, in turn, give rise to topological states that can be described by the bulk-edge correspondence and topological invariants. In particular, topological edge states, which can be quantified by a winding number $\mathcal{W}=\mathcal{Z}/\pi$, appear at the end of a region with non-zero Zak phase $\mathcal{Z}$, where $\mathcal{Z}=0$ or $\pi$ depending on the edge termination [39]. If the lattice terminates with the weak coupling $c_1$, see the upper configuration of Fig. 1b, the winding number is one and the lattice supports one topological edge state. On the contrary, if the lattice terminates with the strong coupling $c_2$, the winding number is zero and the structure does not support an edge state, as it is displayed in the lower configuration of Fig. 1b. The topological protection of these states is directly related with the existence of internal symmetries in the system. Specifically, the chiral symmetry given by $\Gamma\mathcal{H}\Gamma^\dagger=-\mathcal{H}$, entails that the energy spectrum of the system is symmetric around zero, guaranteeing that all the states with positive energy have a counterpart with the same negative energy, with the exception of the zero energy states, which are topologically protected (See Supplementary S2 for more details).

Discrete SUSY transformations applied to the Hamiltonian can be expressed in terms of a transformation matrix $V$ as:
$$V\mathcal{H}_m^{(1)}V^{-1} = VQRV^{-1} = RQ = \mathcal{H}_m^{(2)}, \qquad (3)$$
where $V=Q^{-1}$. If both $\mathcal{H}_m^{(1)}$ and $V$ possess some symmetry, e.g. chiral symmetry satisfying the anti-commutator relation $\{\mathcal{H}_m^{(1)}, \Gamma\}=\{V, \Gamma\}=0$, then this symmetry is transferred to $\mathcal{H}_m^{(2)}$:
$$\mathcal{H}_m^{(2)} = V\mathcal{H}_m^{(1)}V^{-1} = -V\Gamma\mathcal{H}_m^{(1)}\Gamma^\dagger V^{-1} = -\Gamma\mathcal{H}_m^{(2)}\Gamma^\dagger, \qquad (4)$$

On the other hand, if the transformation matrix *V* does not obey this symmetry, it will not be reproduced in the superpartner Hamiltonian $\mathcal{H}_m^{(2)}$ either. Exploiting this connection between symmetry constraints of SUSY transformations and symmetries of the system, superpartner Hamiltonians with modified topological properties can be engineered. To elucidate this, a SSH-type lattice composed of an even number *N* of waveguides, supporting two topologically protected edge states, is considered as the starting point (see Fig. 2b and 2e). As a proof of concept, two distinct superpartner structures are investigated: (i) the superpartner SP$_{N/2}$, obtained by removing the eigenvalue $\lambda_{N/2}$ corresponding to a topological edge state (see Fig. 2c and 2f), and (ii) the superpartner SP$_1$, obtained by removing the eigenvalue $\lambda_1$ corresponding to a bulk state (see Fig. 2a and 2d). Note that, due to the symmetry of the eigenvalue spectrum, equivalent results would be obtained by removing $\lambda_{N/2+1}$ and $\lambda_N$, respectively. Subsequently, the degree of protection of the superpartner topological states is probed analytically with respect to their symmetries, as well as by gauging their robustness against chiral disorder [40].

**Supersymmetric topological photonic structures**

Figure 2c shows the eigenvalue spectrum of the SP$_{N/2}$ lattice obtained by removing the eigenvalue $\lambda_{N/2}$ corresponding to an edge state of the SSH structure. Since it is a zero energy eigenvalue, the diagonal elements of the superpartner Hamiltonians $\mathcal{H}_{N/2}^{(1)}$ and $\mathcal{H}_{N/2}^{(2)}$ remain zero. Thus, the superpartner lattice is composed of waveguides with zero detuning (see Supplementary S1 for an extended discussion). Here, the transformation matrix possesses chiral symmetry, which is transferred to the superpartner Hamiltonian $\mathcal{H}_{N/2}^{(2)}$ that satisfies $\Gamma \mathcal{H}_{N/2}^{(2)} \Gamma^\dagger = -\mathcal{H}_{N/2}^{(2)}$. Therefore, the symmetries of the system are preserved and the topological properties of the remaining zero-energy eigenstate remain intact. By applying SUSY transformations, two different superpartner lattices supporting one topological state $\Psi_{N/2+1}$ can be obtained. One supporting an interface state, as displayed in Fig. 2f, and the other supporting an edge state, mostly maintaining the form of Fig. 2e with the last waveguide removed. For the interface state solution, the SP$_{N/2}$ structure resembles two SSH chains with different termination at the interface and strong coupling at the outer edges. The topologically protected interface state, whose position in the lattice can be controlled by changing the dimerization ratio, is located between the two SSH lattices and decays exponentially into the bulk. The existence of this interface state is experimentally verified, as discussed in detail in the next section, and its robustness against disorder maintaining the underlying symmetry of the lattice is numerically proved. In particular, by introducing chiral disorder, the deviation of the eigenvalue $\lambda_{N/2+1}$ is proved to be zero, while the eigenstate shape is slightly modified although it remains localized at the interface (see Supplementary S2). As expected, non-chiral disorder destroys the topological protection and leads to notable modifications to the eigenvalues. For the edge state solution, the $SP_{N/2}$ structure resembles the SSH model with interchanged couplings and $N$-1 waveguides, except for a localized deviation in the couplings with respect to $c_1$ and $c_2$ near the leading edge. Here, the SUSY transformation constitutes a topological phase transition in the sense that the couplings are interchanged and one waveguide removed, thus, one of the edge states is annihilated. As before, the remaining edge state is topologically protected and robust against chiral disorder. Note that, by applying another SUSY transformation removing the remaining zero-energy eigenvalue, the system becomes topologically trivial. To sum up, by annihilating zero-energy eigenvalues, SUSY transformations introduce topological phase transitions, leading to the creation, displacement, and destruction of topological states.

Let us now consider the $SP_1$ lattice, obtained by removing the eigenvalue $\lambda_1$ corresponding to a bulk state of the SSH structure, as it is displayed in Fig. 2a. Considering that the removal of any bulk state of the system per definition breaks the inversion symmetry of the eigenvalue spectrum, one would expect that the topological protection of the edge states is necessarily destroyed. Nevertheless, the chiral symmetry of the system is partially respected by the SUSY transformation, preserving the topological protection of one edge state. This can be explained by separating the Hamiltonian $\mathcal{H}_1^{(2)}$ into $\mathcal{H}_{1L}^{(2)}$ and $\mathcal{H}_{1R}^{(2)}$, corresponding to the left and right parts of the lattice, respectively. The chiral symmetry of $\mathcal{H}_{1R}^{(2)}$ is preserved, satisfying $\Gamma\mathcal{H}_{1R}^{(2)}\Gamma^\dagger=-\mathcal{H}_{1R}^{(2)}$ and, thus, the topological protection of the right edge-state is maintained. On the contrary, the chiral symmetry of $\mathcal{H}_{1L}^{(2)}$ is destroyed by the appearance of nonzero diagonal elements, which take away the symmetry protection of the left edge state. However, the state remains localized at the edge due to the high detuning between waveguides. The $SP_1$ lattice exhibits an exponentially decaying detuning on the left side of the lattice, while still resembling the SSH model towards the right part of the lattice (see Supplementary Materials S3 for more details). The existence of both edge states and their different origins is experimentally verified, as discussed in the next section. Also, the stability of the edge states eigenvalues in the spectrum is numerically checked by introducing chiral disorder. Specifically, for the right edge state the deviation of the eigenvalue $\lambda_{N/2+1}$ tends to zero as $N$ increases, whereas for the left edge state, the deviation of the eigenvalue $\lambda_{N/2}$ is not affected by the size of the system and increases linearly with the amount of disorder (see Supplementary Materials S2 for more details). Note that by applying another SUSY transformation removing $\lambda_N$, the inversion symmetry of the system is reestablished, and the topological protection of the left edge state can be restored. Moreover, by removing higher-order bulk states only from one side of the spectrum, the detuned region can be extended across the lattice to facilitate an enhanced interaction with the right edge state. Finally, by applying multiple SUSY transformations symmetrically, gaps can be carved out of the eigenvalue spectrum while preserving the topological protection of the zero-energy states. Here, in short, we have transformed a lattice supporting two topologically protected edge states to a phase-matched lattice supporting one topologically protected edge state, and one that has lost its topological protection and has become sensitive to the underlying disorder.

**Experimental verification**

In order to experimentally corroborate the previous theoretical findings, we employ the femtosecond direct laser-writing technology to inscribe waveguide arrays in fused silica (See Methods and Supplementary S3). Specifically, we exploit its ability to independently tune the coupling and detuning by changing the separation between waveguides and the inscription velocity, respectively [22]. To this aim, four different samples are fabricated: (i) the original SSH lattice described by $\mathcal{H}$, (ii) the superpartner $SP_{N/2}$ lattice described by $\mathcal{H}_{N/2}^{(2)}$, (iii) the superpartner $SP_1$ lattice described by $\mathcal{H}_1^{(2)}$ and, (iv) the SSH lattice weakly coupled to the $SP_1$ lattice. By launching single site excitations, light evolution of the different states along the different structures can be measured by means of waveguide fluorescence microscopy [14], and output pattern intensities can be extracted. Furthermore, by using a white light source, the wavelength of the injected light can be continuously tuned to evaluate the robustness and different origins of the edge states. Finally, by placing the SSH lattice in close proximity to the $SP_1$ lattice, evanescent coupling can be introduced between the topological edge state in the former, and the non-topological edge state in the latter. The contrast of the resulting sinusoidal intensity oscillations serves as direct indicator for any detuning between them, or the predicted absence thereof.

The first step to verify the previous theoretical predictions is to prove the existence of the topological edge states of the SSH lattice. To this end, we excite the right edge state by injecting light into the $N^{th}$ waveguide, as depicted in Fig. 2h. As a single-site excitation is made, and the theoretically expected edge state is exponentially localized within the waveguides $N$, $N-2$ and $N-4$, as it is illustrated in Fig. 2e, other bulk states of the system are also excited and the injected intensity slightly spreads along the propagation direction. However, one can clearly observe how the output measured intensity distribution is in accordance with the predicted mode profile, showing the expected SSH edge state. Since the SSH lattice is symmetric, a mirrored propagation image would be obtained by injecting light into the first waveguide, exciting the left topological edge state. Note that, the confinement of this edge state scales with the difference between the coupling coefficients $c_1$ and $c_2$. The next step is to demonstrate the presence of the interface state of the $SP_{N/2}$ lattice. Although the expected theoretical interface state spans approximately five odd waveguides, as depicted in Fig. 2f, it is nevertheless populated by a single site excitation at the interface waveguide, as displayed in Fig. 2i. Moreover, as can be observed from the output intensity pattern, most of the light is localized at the interface waveguide itself. Note that, for the trivial solution corresponding to the $SP_{N/2}$ structure supporting only one edge state, light evolution and output intensity would resemble the previously obtained for the SSH lattice. The next stage is to prove the existence of the non-topological edge state of the $SP_1$ lattice, which has lost its topological protection due to the breaking of chiral symmetry of one part of the system. To do that, the first waveguide of the $SP_1$ lattice is excited, as it is displayed in Fig. 2g. While the localization is still visible, it may be noted that the intensity distribution is clearly different from the topological state, as depicted in Fig. 2d. Since this edge state is solely mediated by the detuning, it is less robust against perturbations than the topological state, as we numerically verified in Supplementary S2. Furthermore, a strong indication to this reasoning can be seen when we excite both edge states tuning the wavelength continuously from 500 nm to 720 nm. The experimental results obtained for the propagation of the different states are in good agreement with the tight-binding simulations (See Supplementary S4 for a detailed discussion).

To verify the different origin of the edge states of the SP$_1$ lattice, we excite both edges with different wavelengths and observe the output intensities after 10 cm of propagation, as can be seen in Fig. 3. Experimentally, this is achieved by using a white light source combined with a narrow wavelength filter, as discussed in detail in the Methods section. The first observation is that, although their different topological nature, both edge states remain localized at the corresponding edges. However, since the non-topological edge state is supported by the detuning, its degree of localization strongly decreases towards longer wavelengths (See Fig. 3a). This occurs because at longer wavelengths, the coupling substantially increases while the detuning decreases, thus the former becomes the dominant term and the confinement of the edge state is reduced. On the contrary, it gets fully localized into a single waveguide for shorter wavelengths, where the detuning is the dominant term. The confirmation that the existence of this edge state is due to the detuning, is a strong indication for less robustness, since it does not have a topological origin. On the other hand, as shown in Fig. 3b, the topological state strictly maintains its characteristic staggered intensity structure across the investigated spectral range. Note that the slight delocalization at short wavelengths occurs as both couplings decrease and their absolute difference $|c_1-c_2|$, which is related with the edge state confinement, becomes too small to strongly confine the state at the edge.

So far, we have proved the existence of the different topological states, as well as the different origin of the edge states of the SP$_1$ lattice. The last step is to verify that the non-topological edge state indeed does possess a zero-energy eigenvalue, as expected from SUSY transformation. To this aim, we weakly couple the non-topological edge state with the topological state, as displayed in Fig. 4c. Here, if the two states have the same energy, one would expect their coupling with a full exchange of power. On the contrary, if the two states have different energies, one would expect only a partial exchange of power. In Fig. 4a, we show the evolution of the power along the propagation direction for the edge waveguides supporting the topological and non-topological edge states. In both cases, a full exchange of energy between edge states can be observed. Furthermore, the intensity oscillations are in good agreement with the tight binding simulations, shown by the dashed lines of Fig. 4a. The full oscillation pattern between edge states, both experimental and simulated, can be observed in Supplementary S4.

Finally, an additional check that both edge states share the same energy is made exciting them with different wavelengths, as displayed in Fig. 4b. By increasing the wavelength, the coupling increases, leading to a reduced effective length scale of the chip. Looking at the output intensities, the full exchange of intensity between waveguides can be observed, confirming that both superpartner share the same eigenvalue spectrum.

**Discussion**

In our work we studied the interplay between topological non-trivial systems and SUSY transformations. For this, we picked one of the most prominent models for illustrating topological physics, the SSH model, and demonstrated how topological phase transitions can be induced by SUSY transformation. While this topological transition may suspend the topological protection of a state, it can readily be reestablished by applying another SUSY transformation. We exemplified this by transforming a lattice supporting two topological edge states to a lattice supporting (i) one topological edge or interface state, and, (ii) one topologically and one non-topological edge states. We experimentally demonstrated those theoretical findings implementing the superpartner structures using femtosecond laser written waveguides.

Clearly, SUSY techniques constitute a powerful tool to design structures with desirable topological properties, which can be extended to higher dimensions and chiral edge states in future works. Moreover, iterative SUSY transformations could serve to remove any number of states from the system and reduce its overall size while preserving the desired part of the spectrum and the system's topological properties. Finally, note that discrete SUSY-like transformations can be extended to any platform allowing independent control of the coupling and detuning of the sites.

# Methods

**Experimental Design**

Our experiments were conducted in femtosecond laser written photonic lattices, where the above mentioned structures are fabricated and characterized as described below.

**Fabrication of the structures**

The waveguides were fabricated in 10 cm fused-silica glass (Corning 7980) samples by using the femtosecond laser writing method [14]. The laser system consists of a Coherent RegA 9000 amplifier seeded with a Coherent Vitara S Ti:Sa laser with an energy of 250 nJ at 800 nm, 100 kHz repetition rate, and a pulse width of approximately 130 fs. By moving the sample with speeds between 91 to 103 mm min$^{-1}$, the refractive index change at the focal point was around $7 \times 10^{-4}$. The created waveguides exhibit a mode field diameter of about 10.4 µm × 8 µm at 633 nm. The propagation losses and birefringence are estimated to be 0.2 dB cm$^{-1}$ and $1 \times 10^{-7}$, respectively.

**Characterization of the structures**

In order to probe the propagation, the samples were illuminated with light from a Helium-Neon laser at 633 nm (Melles-Griot). The single lattice sites were excited with a 10× microscope objective (0.25NA). In turn, the color centers that formed during the fabrication process, enable a direct observation of the propagation dynamics by using fluorescence microscopy [14]. The recorded images were post processed to reduce noise, distortions and the influence of background light.

The intensities at the output facet at different wavelength were measured by using a white light source (NKT SuperK EXTREME) combined with a narrow wavelength filter (Photon ETC LLTF-SR-VIS-HP8). The light is then coupled into a single lattice site of the sample with a 10× microscope objective (0.25NA) and the resulting light at the output facet of the sample is imaged onto a CCD camera (BASLER Aviator) with another 10× microscope objective. The recorded images were post-processed to reduce noise and subsequently integrated over a strip along the direction perpendicular to the lattice orientation for each wavelength. The resulting intensity distribution for the different wavelengths are then normalized to the maximum value to increase the visibility.

## Acknowledgements

G.Q., J.M., and V.A. acknowledge financial support by Spanish Ministry of Science, Innovation and Universities MICINN (Contract No. FIS2017-86530-P) and Generalitat de Catalunya (Contract No. SGR2017-1646). G. Q. also acknowledges the financial support of German Academic Exchange Service (DAAD). A.S. thanks the Deutsche Forschungsgemeinschaft for funding this research (grants BL 574/13-1, SZ 276/15-1, SZ 276/20-1). The authors would like to thank C. Otto for preparing the highly-quality fused silica samples used in all our experiments.

## Author contributions

G.Q., M.K., and M.H. developed the theory. M.K., L.M., and G. Q. fabricated the samples and performed the measurements. M.H., J.M., V.A., and A.S. supervised the project. All authors discussed the results and co-wrote the paper.

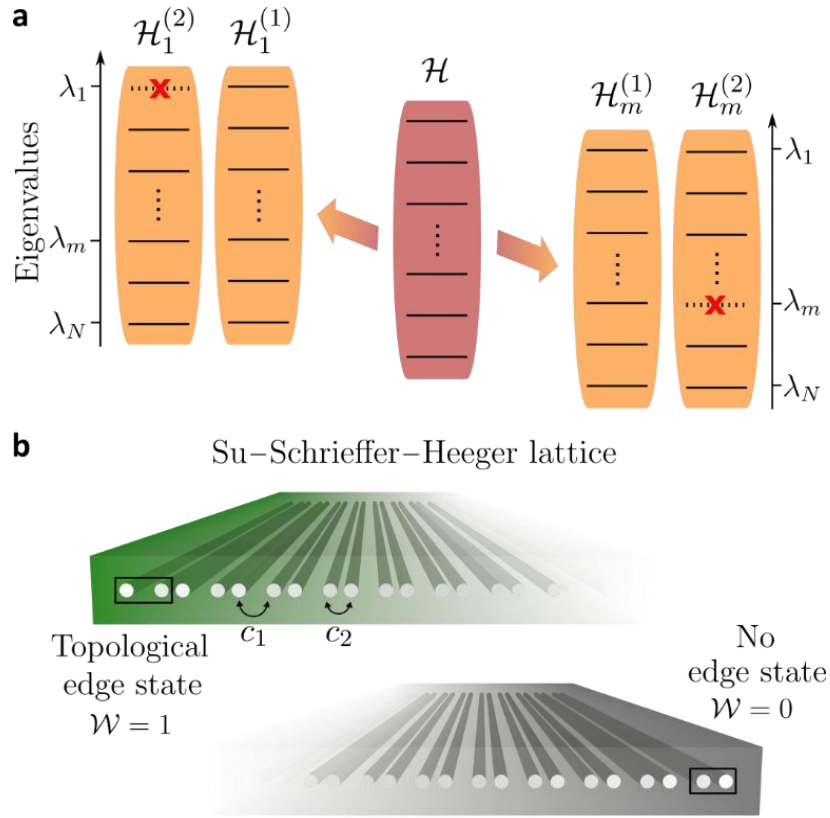

**Fig. 1. Supersymmetric transformations and SSH model. a,** Schematic representation of the eigenvalue spectrum of the Hamiltonian $\mathcal{H}$ and two sets of superpartner Hamiltonians $\{\mathcal{H}_1^{(1)}, \mathcal{H}_1^{(2)}\}$ and $\{\mathcal{H}_m^{(1)}, \mathcal{H}_m^{(2)}\}$, obtained by removing the eigenvalues $\lambda_1$ and $\lambda_m$ using SUSY transformations, respectively. **b,** Representation of a SSH-like lattice implemented using optical waveguides, which are evanescently coupled with alternating couplings $c_1$ and $c_2$ ($c_1 < c_2$). Depending on the termination of the lattice, the structure has Winding number $\mathcal{W}=1$ and supports a topological edge state on that edge (upper configuration) or $\mathcal{W}=0$ and does not support an edge state on that edge (lower configuration).

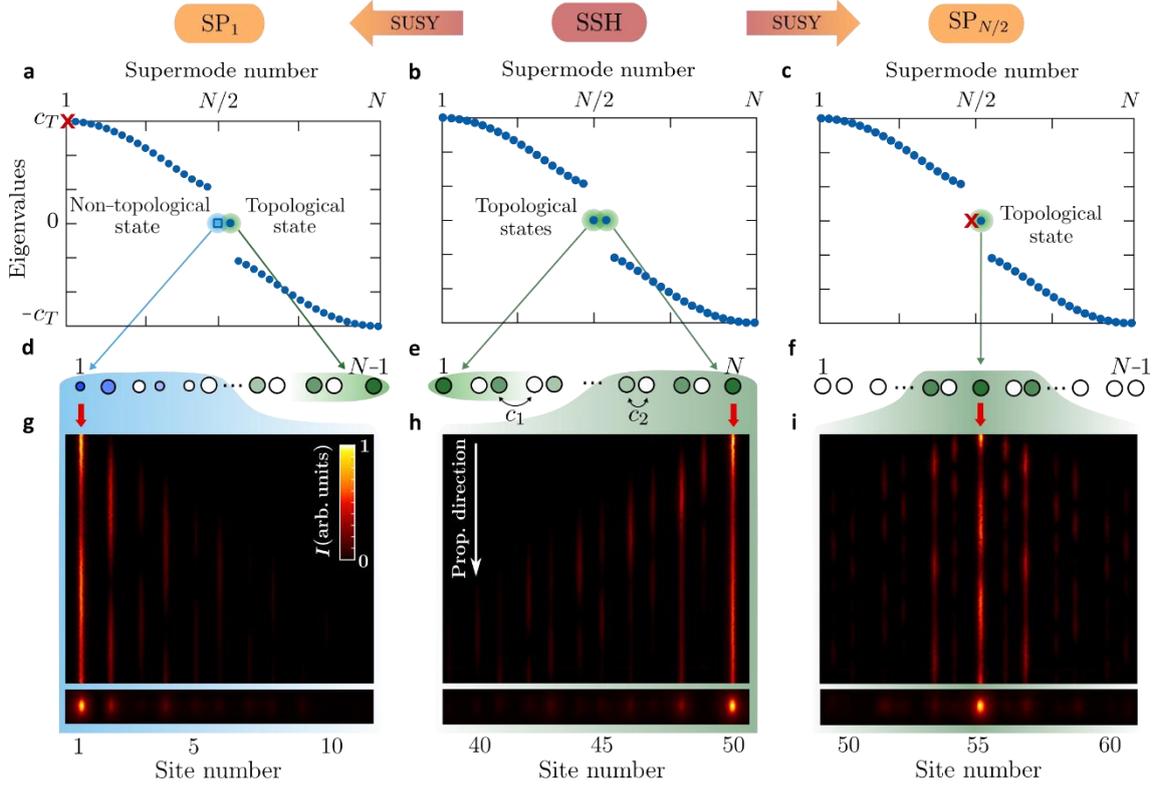

**Fig. 2. Supersymmetric photonic topological structures.** Upper row: eigenvalue spectrum of the **b,** SSH and two Superpartners (SP$_m$) lattices, obtained by removing **a,** a bulk state ($m = 1$) and **c,** an edge-state ($m = N/2$), respectively. The energy gap is $2|c_1-c_2|$ and $c_T = c_1 + c_2$. Central row: sketch of the **d,** SP$_1$, **e,** SSH and **f,** SP$_{N/2}$ lattices. Detuning (coupling) is indicated by the size (spacing) of (between) the circles. The intensity of the color inside each waveguide is proportional to the amplitude of the state. Lower row: experimentally observed light evolution along the propagation direction (top) and output intensities (bottom) for the **g,** non-topological edge state, **h,** topological edge state and **i,** topological interface state. The total length of the sample is $L$=10 cm and the wavelength used to excite the waveguides is $\lambda$ =633 nm. The SSH, SP$_1$ and SP$_{N/2}$ lattices are composed of $N$=50, $N$=49 and $N$=109 waveguides, respectively.

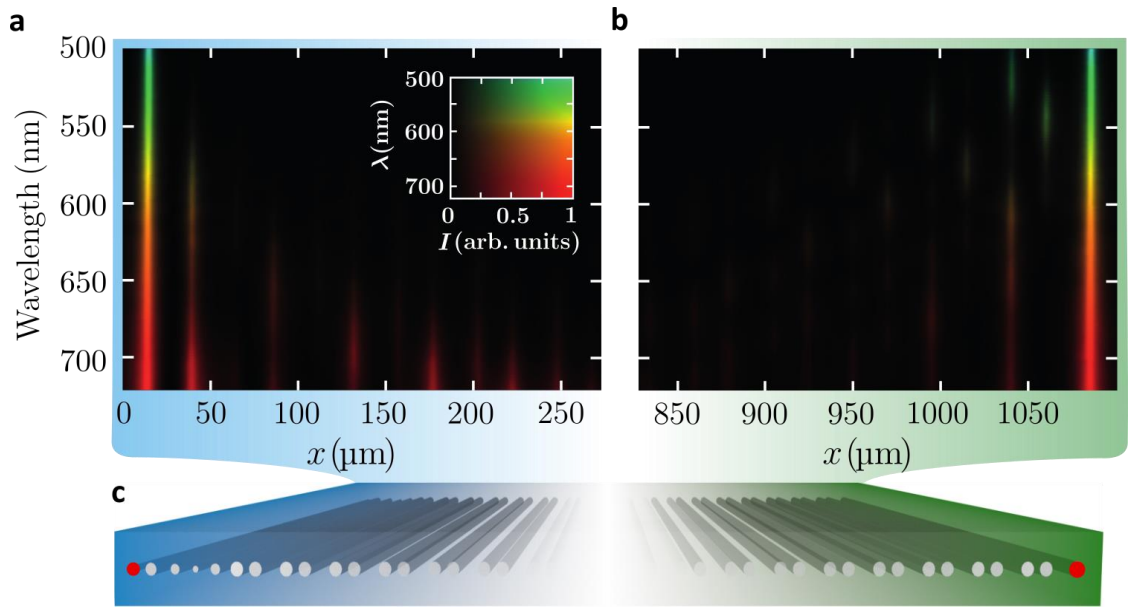

**Fig. 3. Robustness of the edge states.** Experimentally observed output intensities for different wavelengths (500 nm ≤ $\lambda$ ≤ 720 nm) obtained by exciting the **a,** non-topological and **b,** the topological edge states of the $SP_1$ lattice, schematically represented in **c**. The red dots indicate the excited waveguides. The relation between the wavelength used and its intensity measured at the output is represented in the inset. The total length of the sample is *L*=10 cm and the $SP_1$ lattice is composed of *N*=49 waveguides.

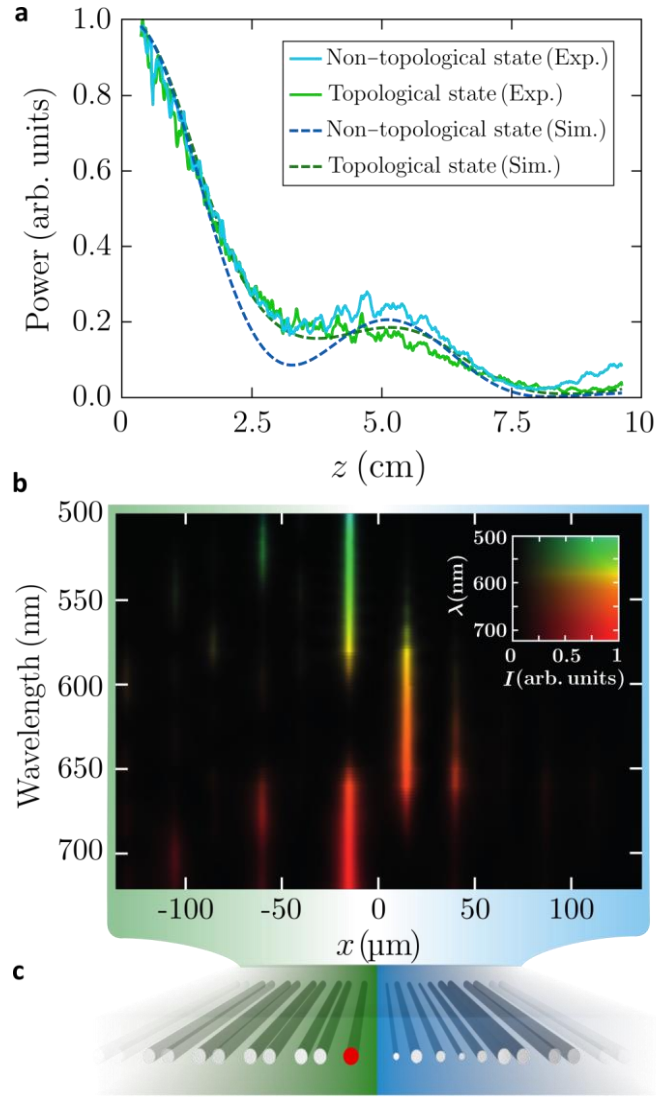

**Fig. 4. Coupling topological and non-topological edge states. a**, Power oscillations when the waveguides supporting the topological and the non-topological edge states of the coupled structure, schematically displayed in **c**, are excited. The solid lines correspond to the experimental results while the dashed lines correspond to the tight-binding numerical simulations. **b**, Experimentally observed output intensities for different wavelengths obtained by exciting the topological edge states of the coupled lattice, indicated with the red dot. **c**, Schematic representation of the SSH lattice (left) weakly coupled to the $SP_1$ lattice (right). The relation between the wavelength used and its intensity measured at the output is represented in the inset. The total length of the sample is $L$=10 cm, the wavelength range used to excite the waveguides is 500 nm $\leq \lambda \leq$ 720 nm and the lattice is composed of $N$=99 waveguides.

## - Supplementary material -

### Section S1. SUSY transformations

In the quantum-mechanical formalism [1], SUSY connects an operator $\mathcal{H}^{(1)}=A^{\dagger}A$, which can be decomposed in terms of an operator $A$ and its Hermitian adjoint $A^{\dagger}$, with $\mathcal{H}^{(2)}=AA^{\dagger}$. From the eigenvalue equation:

$$\mathcal{H}^{(1)}\Psi_s^{(1)} = \lambda_s^{(1)}\Psi_s^{(1)}, \qquad (S1)$$

where $\lambda_s^{(1)}$ is the eigenvalue and $\Psi_s^{(1)}$ is the eigenstate of state $s$, one can derive:

$$A\mathcal{H}^{(1)}\Psi_s^{(1)} = A(A^{\dagger}A)\Psi_s^{(1)} = \mathcal{H}^{(2)}\left(A\Psi_s^{(1)}\right) = \lambda_s^{(1)}\left(A\Psi_s^{(1)}\right), \qquad (S2)$$

obtaining that $A\Psi_s^{(1)}$ is an eigenstate of $\mathcal{H}^{(2)}$ with eigenvalue $\lambda_s^{(1)}$, establishing SUSY isospectrality. For unbroken SUSY, the ground state of $\mathcal{H}^{(1)}$ is annihilated by $A\Psi_s^{(1)}$ and removed from the spectrum of $\mathcal{H}^{(2)}$. SUSY-like transformations can be extended to discrete systems [2], by means of symmetric and asymmetric methods such as the Cholesky algorithm and the *QR* factorization, respectively [3]. The latter method allows to remove any eigenvalue of the spectrum without resorting to non-Hermitian configurations. In general, a *QR* factorization of a matrix $B \in \mathbb{R}^{m\times n}$ ($m \geq n$) is a decomposition into $B=QR$, where $Q \in \mathbb{R}^{m\times m}$ is an orthogonal matrix and $R \in \mathbb{R}^{m\times n}$ is an upper triangular matrix [3]. Considering a discrete SSH-type Hamiltonian $\mathcal{H}$, describing a system of *N* identical evanescently coupled waveguides:

$$\mathcal{H} = \begin{pmatrix} \Delta\beta & c_1 & 0 & \cdots & 0 \\ c_1 & \Delta\beta & c_2 & \ddots & \vdots \\ 0 & c_2 & \ddots & \ddots & 0 \\ \vdots & \ddots & \ddots & \ddots & c_1 \\ 0 & \cdots & 0 & c_1 & \Delta\beta \end{pmatrix} \qquad (S3)$$

where $c_1$ and $c_2$ are the coupling strengths and $\Delta\beta$ the detuning, the superpartner Hamiltonians obtained using the *QR* factorizations are:

$$\mathcal{H}_m^{(1)} = \mathcal{H} - \lambda_m I = QR, \qquad \mathcal{H}_m^{(2)} = RQ, \qquad (S4)$$

where $\mathcal{H}_m^{(2)}$ represents a lattice with *N* waveguides and the same eigenvalues as $\mathcal{H}_m^{(1)}$ except for $\lambda_m$. This discrete SUSY-like transformation is exemplified for *N*=6 in Fig. S1 a-d, in which starting with the SSH model we eliminate $\lambda_3$ from the eigenvalue spectrum. In particular, to perform the *QR* factorization, we use the Givens Rotation method that is numerically stable, thus, suitable for application on large lattices [3]. This method is based on applying rotations given by matrices $G_j = g_j \otimes I$ to $\mathcal{H}_m^{(1)}$, forming:

$$R = \prod_{j=1}^{N-1} G_j \mathcal{H}_m^{(1)}, \quad Q = \prod_{j=1}^{N-1} G_j^T. \tag{S5}$$

The rotations $g_j \in \mathbb{R}^{2\times 2}$ introduce zeroes at the subdiagonal elements $a_{j+1,j}$ of $\mathcal{H}$ as:

$$g_j \begin{pmatrix} a_{j,j} \\ a_{j+1,1} \end{pmatrix} = \begin{pmatrix} t_j & s_j \\ -s_j & t_j \end{pmatrix} \begin{pmatrix} a_{j,j} \\ a_{j+1,1} \end{pmatrix} = \begin{pmatrix} r_j \\ 0 \end{pmatrix}, \tag{S6}$$

where $t_j = a_{j,j}/r_j$, $s_j = a_{j+1,j}/r_j$ and $r_j = [a_{j,j}^2 + a_{j+1,j}^2]^{1/2}$, which can be rewritten in terms of the corresponding Pauli matrices as:

$$g_j = t_j \sigma_0 + i s_j \sigma_y. \tag{S7}$$

Recall that the Pauli matrices are given by:

$$\sigma_0 = \begin{pmatrix} 1 & 0 \\ 0 & 1 \end{pmatrix}, \quad \sigma_x = \begin{pmatrix} 0 & 1 \\ 1 & 0 \end{pmatrix}, \quad \sigma_y = \begin{pmatrix} 0 & -i \\ i & 0 \end{pmatrix}, \quad \sigma_z = \begin{pmatrix} 1 & 0 \\ 0 & -1 \end{pmatrix}. \tag{S8}$$

These rotation functions $g_j$ can be related to the topological transitions. For the SP$_{N/2}$ lattice (Fig. S3i) supporting an interface state, we can observe in Figs. S1f and S1h, how $s_j \to 1$ and $t_j \to 0$ for $j < j_{\text{interface}}$. Thus, $g_j \to i\sigma_y$, interchanging the couplings $c_1$ and $c_2$ of the original SSH lattice. Around $j_{\text{interface}}$, there is a transition to a more complex behavior $g_{j,\text{even}} \to i\sigma_y$, $g_{j,\text{odd}} \to \pm I + i\sigma_y c_2/c_1$, which leads to a recovering of the original SSH configuration. Note that, since $\lambda_{N/2} \sim 10^{-16}$, the detunings appearing at the superpartner lattice are of the same order and does not have any influence on the system. Considering $\lambda_{N/2} = 0$, a more trivial solution is obtained with $g_j \to i\sigma_y$, constituting a topological phase transition, giving a new structure resembling the SSH model with interchanged couplings and $N$-1 waveguides. For the SP$_1$ lattice (Fig. S3g), we can observe in Figs. S1e and S1g, how $g_j \to i\sigma_y$ towards the right part of the lattice, where the superpartner still resembles the SSH model, while $g_j \sim t_j \sigma_0$ is the dominant term throughout the left side of the lattice, inducing a small deviation in the couplings with respect to the original SSH lattice and introducing an exponentially decaying detuning.

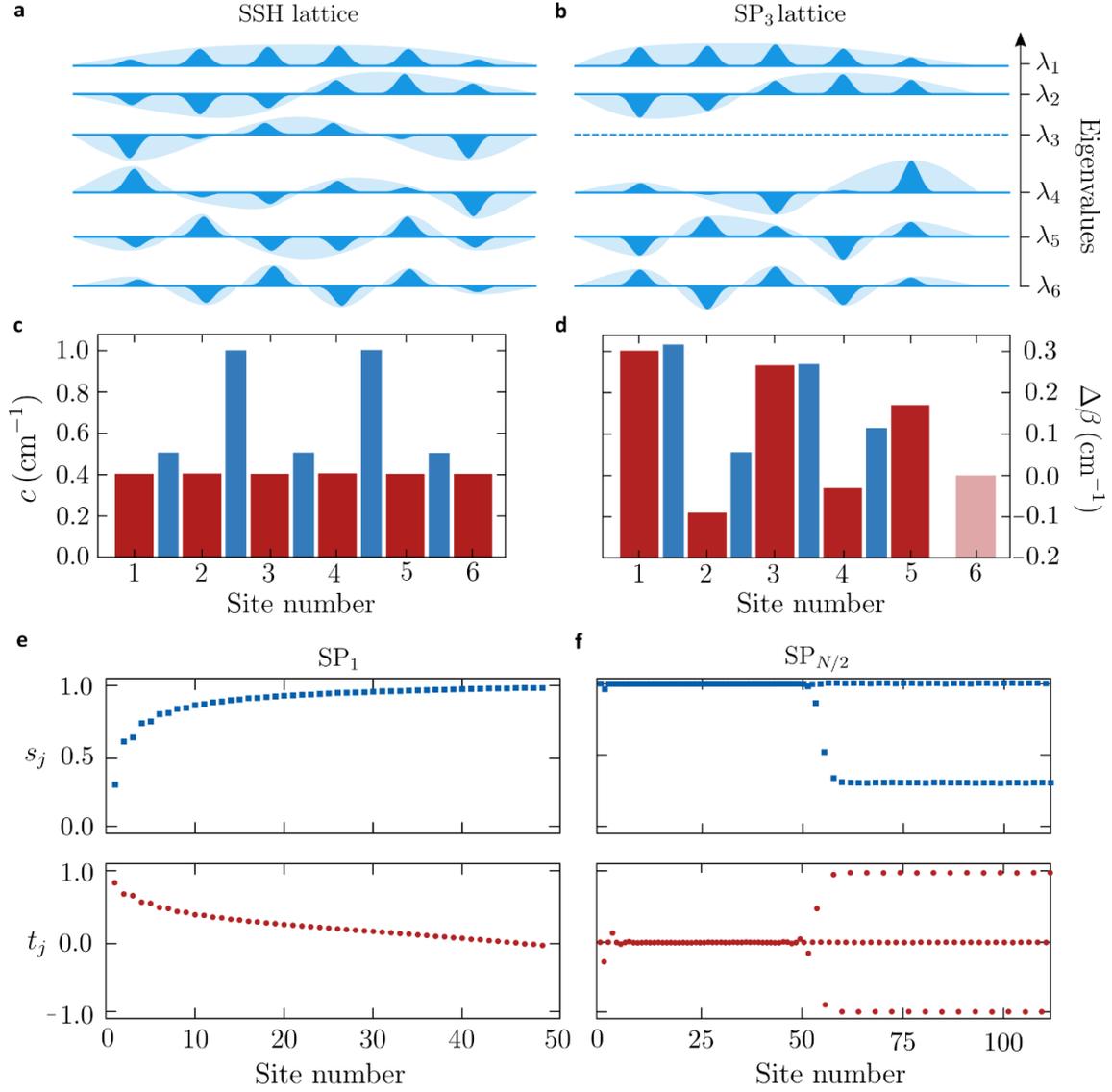

**Fig. S1. Discrete Optical supersymmetry.** Eigenvalue spectrum and eigenstate profiles corresponding to **a,** the SSH lattice and **b**, the $SP_3$ lattice, obtained by removing the eigenvalue $\lambda_3$ using discrete SUSY transformations. Discrete representation in terms of the waveguide's detuning $\Delta\beta$ (red bars) and couplings $c$ (blue bars) of the **c,** SSH and **d,** $SP_3$ lattices. The SSH lattice is composed by $N=6$ waveguides, $c_1=0.5$ cm$^{-1}$ and $c_2=1.0$ cm$^{-1}$. Representation of $s_j$ and $t_j$ obtained using the Givens Rotation method to perform the QR factorization for **e,** the $SP_1$ lattice of Fig. S3g and **f,** the $SP_{N/2}$ lattice of Fig. S3i.

## Section S2. Robustness of the topological states

The Hamiltonian $\mathcal{H}$ of the SSH model, given by Eq. (S3), is an $N \times N$ tridiagonal matrix with zero-valued diagonal elements ($\Delta\beta = 0$) and off-diagonal elements alternating between $c_1$ and $c_2$ ($c_1 < c_2$). This system supports two topologically protected zero-energy edge states, which appear at the edge of a region with non-zero Zak phase $\mathcal{Z}$:

$$\mathcal{Z} = i \oint u^*(k) \partial_k u(k) dk, \qquad (S9)$$

where $k$ is the Bloch wavenumber within the first Brillouin zone and $u(k)$ the corresponding eigenvector in $k$-space [4]. The topological protection of these states is closely related with the existence and breaking of the symmetries of the system. In particular, the SSH model can be characterized with two main symmetries (i) the Chiral symmetry and, (ii) the Particle Hole symmetry (PHS). On the one hand, the Chiral or Sublattice symmetry, is defined by the unitary and Hermitian operator $\Gamma$, which anti-commutes with the Hamiltonian $\mathcal{H}$, hence, $\{\mathcal{H}, \Gamma\}=0$. On the other hand, the PHS is defined by an anti-unitary operator P, which also anti-commutes with the Hamiltonian $\mathcal{H}$, having $\{\mathcal{H},P\}=0$. Note that, since chiral and PHS exist, time reversal symmetry also exists for this model. The application of these operators to the Hamiltonian leads to:

$$\Gamma \mathcal{H} \Gamma^\dagger = -\mathcal{H}, \quad P\mathcal{H}P^{-1} = -\mathcal{H}. \qquad (S10)$$

The chiral symmetry of the system is responsible for the topological protection of the zero-energy states. By applying the operator $\Gamma^\dagger$ to the eigenvalue equation $\mathcal{H}\Psi_s = \lambda_s \Psi_s$, one obtains:

$$\mathcal{H}\Gamma^\dagger \Psi_s = -\lambda_s \Gamma^\dagger \Psi_s. \qquad (S11)$$

Therefore, for any state $\Psi_s$ with eigenvalue $\lambda_s$, a symmetric partner with energy $-\lambda_s$ exists, except for the zero-energy eigenvalue. This symmetry of the spectrum leads to the symmetric protection of the zero-energy states, which are robust against disorder maintaining the underlying symmetry of the system [5]. To numerically prove the topological protection of the states, we introduce chiral disorder of the form:

$$\tilde{c}_1^q = c_1 + \Delta c, \quad \tilde{c}_2^q = c_2 - \Delta c, \qquad (S12)$$

where $\Delta c = K \xi_q |c_1 - c_2|$. The disorder is quantified in terms of the disorder strength $K$, the dimerization ratio $|c_1 - c_2|$, and a random number $-1 \leq \xi_q \leq 1$, and affects the couplings of each unit cell $q$, formed by two sites. Note that, the random numbers introduced are different for each unit cell $q$. To prove the robustness, we compute the deviation of the eigenvalues and eigenstates with respect to the case without disorder:

$$\Delta\lambda_s = |\lambda_s - \lambda_s(K=0)|, \quad \Delta\Psi_s = |\Psi_{s,j} - \Psi_{s,j}(K=0)|, \qquad (S13)$$

where $j$ accounts for each site of the lattice. Due to their symmetries, the zero-energy eigenvalues, corresponding to the topological edge states, should be robust against this kind of disorder, while the other states should exhibit an eigenvalue deviation.

For the SP$_1$ lattice of Fig. S3g, by introducing up to 25% of disorder with respect of $|c_1 - c_2|$, we can observe in Fig. S2a how there is no deviation of the eigenvalue $\lambda_{26}$ corresponding to the topological edge state. On the other hand, if we take a look at the deviation of the eigenvalue $\lambda_{25}$ corresponding to the non-topological edge state, we can confirm that the state is not topologically protected and its eigenvalue increases linearly with the amount of disorder. Note that for the topological edge state the deviation of the eigenvalue $\lambda_{26}$ reduces as $N$ increases, whereas the eigenvalue deviation of the non-topological edge state $\Delta\lambda_{25}$ is not affected by the size of the system. Regarding the changes in the eigenstate shapes, shown in Fig. S2b, we can observe how the non-topological edge state suffers more deviations than the topological one. In both cases, the change in the eigenstate shapes are small, and the states remain localized at the corresponding edges of the SP$_1$ lattice.

For the SP$_{N/2}$ lattice of Fig. S3i, it is also numerically shown that by introducing up to 25% of disorder with respect of $|c_1 - c_2|$, there is no deviation of the zero-energy eigenvalue $\lambda_{56}$ corresponding to the topological interface state, as can be seen in Fig. S2a. In this case, although the eigenstate shape is more perturbed than for the topological edges, it still remains localized at the interface. The higher deviation in this case may be produced due to the fact that the interface state spreads along more waveguides than the edge states. Thus, disorder introduced to the system has more impact on the modification of its shape, while it does not affect its eigenvalue which is protected by the symmetries of the system. Finally, if any other kind of disorder not preserving the chiral symmetry of the system is introduced, the topological states are no longer topologically protected and their zero-energy eigenvalues suffer deviations of the same order than the non-topological states.

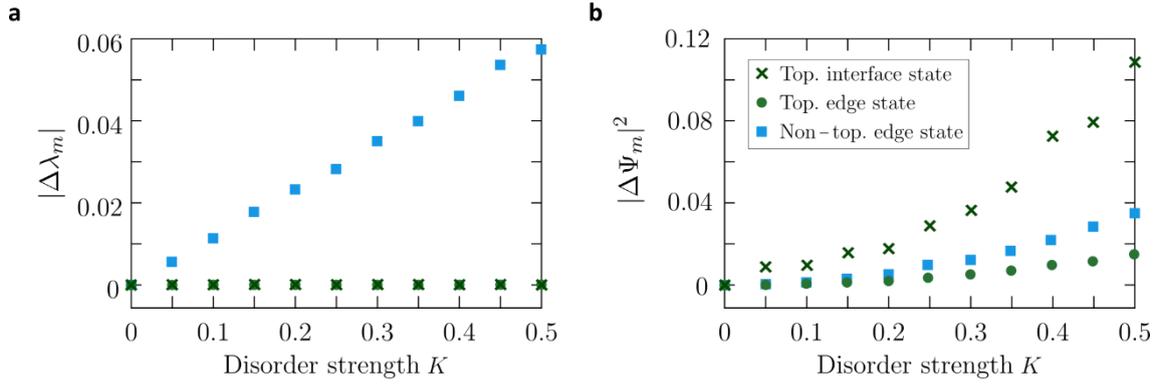

**Fig. S2. Robustness against chiral disorder.** Deviation of the **a,** eigenvalues $\lambda_s$ and **b,** eigenstate shapes $\Psi_s$ with respect to the ones of the original lattice ($K=0$) when chiral disorder is introduced into the system, for: the non-topological edge state $s=25$ (squares), the topological edge state $s=26$ (circles) and the topological interface state $s=56$ (crosses). The total deviation is averaged over 1000 different simulations with different random disorder. All the simulations were carried out using the lattices of Figs. S3 g-i, with the $SP_1$ and the $SP_{N/2}$ having 49 and 110 waveguides, respectively.

## Section S3. Design of the SUSY structures

The experimental implementation of the SUSY structures with femtosecond laser written waveguides is based on the similarity between the Schrödinger and the Helmholtz equations [6]. On the one hand, the Schrödinger equation of quantum mechanics is given by:

$$i\hbar \frac{\partial}{\partial t}\Psi(x,y,t) = -\frac{\hbar^2}{2m}\nabla^2\Psi(x,y,t) + V(x,y,t)\Psi(x,y,t), \qquad (S13)$$

where $\Psi(x,y,t)$ is the wavefunction, $\hbar = h/2\pi$ is the reduced Planck constant and $V(x,y,t)$ is the potential. On the other hand, the paraxial Helmholtz equation is given by:

$$i\lambdabar \frac{\partial}{\partial z}E(x,y,z) = -\frac{\lambdabar^2}{2n_0}\nabla^2 E(x,y,z) - \Delta n(x,y,z)E(x,y,z), \qquad (S14)$$

In turn, wavefunction $\Psi(x,y,t)$ in the Schrödinger equation is replaced by the electric field amplitude $E(x,y,z)$ in the Helmholtz equation, the potential $V(x,y,t)$ is replaced by the refractive index profile $-\Delta n(x,y,z)$, the propagation in time is replaced by the spatial coordinate $z$, which can be monitored by means of fluorescence microscopy, and the reduced Planck constant is replaced by the reduced wavelength $\lambdabar = \lambda/2\pi$. In the tight-binding approximation, Eq. (S14) furthermore simplifies to:

$$i\frac{\partial}{\partial z}\psi_j = \beta_j \psi_j + c_{j,j+1}\psi_{j+1} + c_{j,j-1}\psi_{j-1}, \qquad (S15)$$

where $\psi_j$ is the field amplitude at site $j$, $\beta_j$ the propagation constant and $c_{j,j\pm 1}$ the coupling between adjacent waveguides.

For the implementation of the SUSY structures, both, the coupling as well as the detuning need to be tuned individually. The coupling ($c$) is changed by using different distances ($d$) between waveguides, while the detuning ($\Delta\beta$) is changed by using different writing velocities ($v$). The relation between distance and coupling (writing speed and detuning) is retrieved from directional couplers, by measuring the coupling length and the intensity contrast. The results are plotted in Fig. S3 a-c. The coupling is well fitted by an exponential function of the distance between waveguides, while the detuning depends linearly on the writing speed. The exponential and linear fits are:

$$c(d) = k_1 \exp(-k_2\, d), \quad k_1 = 10.93\ \text{cm}^{-1},\ k_2 = 0.121\ \mu\text{m}^{-1},$$
$$\Delta\beta(v) = k_3\, v + k_4, \quad k_3 = -0.8327\ \text{min cm}^{-2},\ k_4 = 8.376\ \text{cm}^{-1}.$$

Note that the coupling is virtually unaffected by changes in the writing speed, as shown in Fig. S3b. This allows for an independent tuning of both parameters for a wide parameter range. The couplings and detunings used in the fabrication process are displayed in Fig. S3 g-i. The edge and interface eigenstates are displayed in Fig. S3 d-f. Note that, the $SP_1$ ($SP_{N/2}$) structure is obtained taking an original SSH lattice with $c_1$=0.5 cm$^{-1}$ and $c_2$=1.0cm$^{-1}$ ($c_2$=1.8cm$^{-1}$). Although a bigger contrast between the couplings would be better to have more localized edge states, for the $SP_1$ lattice we had the experimental restriction to adjust the detuning below 2 cm$^{-1}$ in order to guarantee that the coupling stays constant when changing the detuning. This could not be fulfilled when using higher couplings for the SSH lattice.

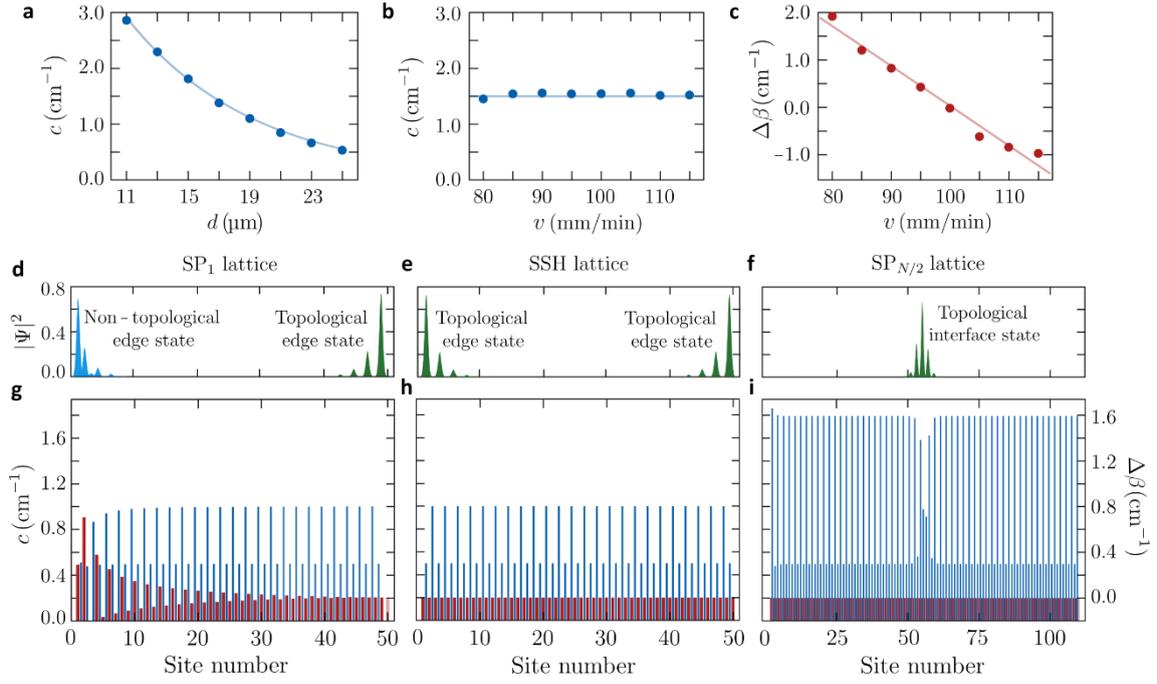

**Fig. S3. Experimental implementation of the SUSY lattices.** Coupling coefficient $c$ dependence with respect to **a,** the waveguide separation $d$, calibrated using pairs of evanescently coupled waveguides with different separation, and, **b**, the writing speed $v$, calibrated using pairs of evanescently coupled waveguides written with different velocities. **c,** Detuning $\Delta\beta$ dependence with respect to the writing velocity. The dots correspond to the experimentally obtained values while the lines are the **a**, exponential and (**b-c**) linear fits. The experimental error associated to the couplings and detunings is ±0.05 cm$^{-1}$. Eigenstate amplitudes of the edge and interface states for the **d**, SP$_1$, **e**, SSH and, **f**, SP$_{N/2}$ lattices. Discrete representation in terms of the detunings $\Delta\beta$ (red bars) and couplings $c$ (blue bars) of the **g**, SP$_1$, **h**, SSH, and **i**, SP$_{N/2}$ lattices. The original SSH lattice to construct the SP$_1$ (SP$_{N/2}$) lattice is composed by $N$=110 ($N$=50) waveguides, $c_1$=0.5 cm$^{-1}$ and $c_2$=1.0 cm$^{-1}$ ($c_2$=1.8 cm$^{-1}$).

## Section S4. Experimental vs simulated propagation images

The intensity distribution is extracted by means of fluorescence microscopy [6], as described in the Methods section. The resulting patterns are compared to tight-binding simulations, obtained by numerically integrating Eq. (S15). The results are displayed in Fig. S4, showing an overall good agreement between the experiments and the simulations, confirming the validity of the theoretical description. Note that a quantitative discrepancy in Fig. S4c, due to locally increased coupling caused by tightly spaced waveguides at the interface, can be observed. However, the qualitative behavior is reproduced.

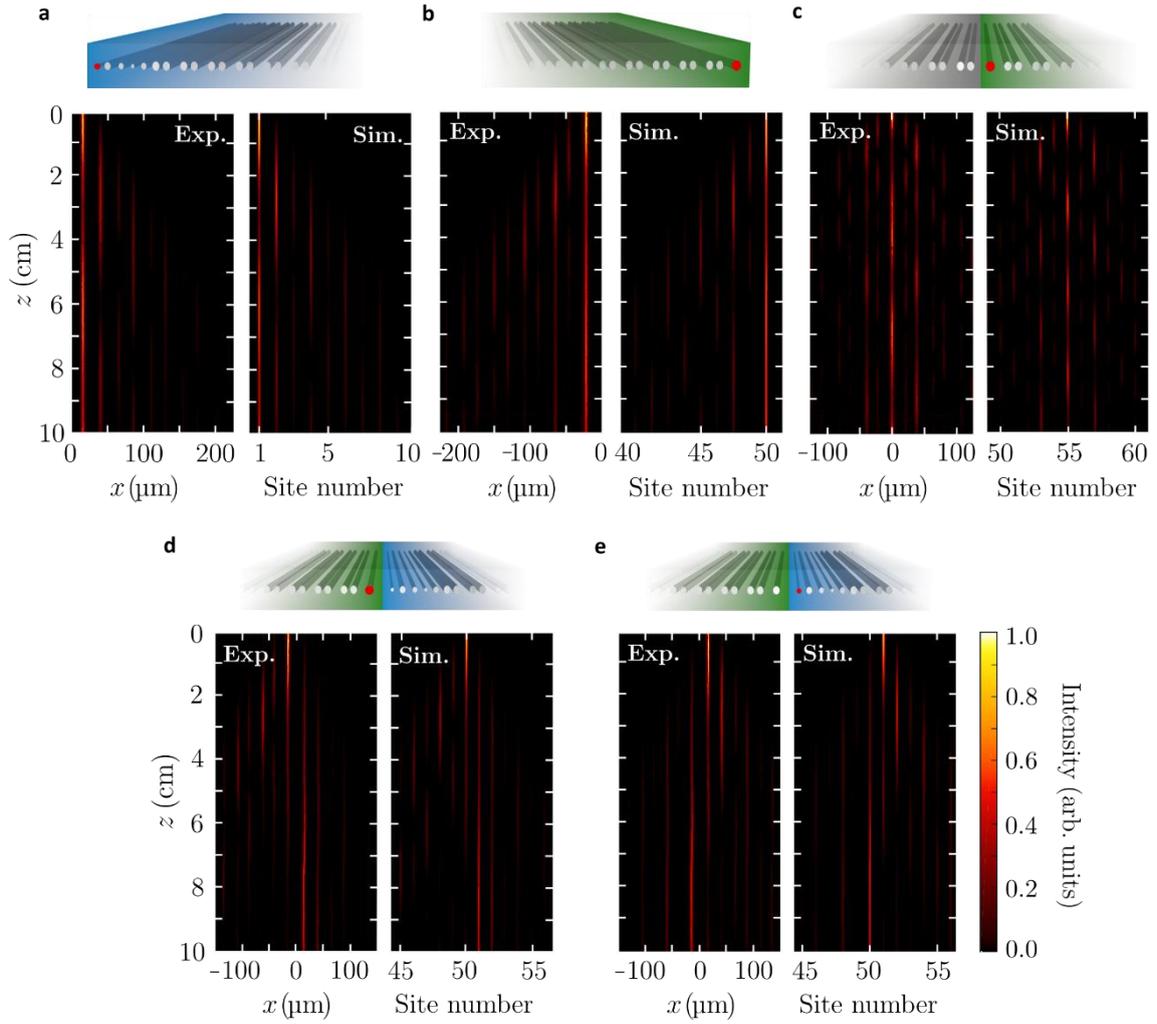

**Fig. S4 Experimental vs numerical simulations.** Intensity distribution, extracted by means of fluorescence microscopy (left), is plotted together with tight-binding binding simulations (right), for the **a,** $SP_1$ lattice, **b**, SSH lattice, **c**, $SP_{N/2}$ lattice, **d-e**, coupled SSH and $SP_1$ lattices. The corresponding lattices are shown schematically above the propagation pictures. The red dots of each structure indicate the excited waveguide, corresponding to the **a**, and **e**, non-topological edge state, **b**, and **d**, topological edge state, and **c**, topological interface state. All the simulations have a correction of $\Delta c =-0.05$ cm$^{-1}$ with respect to the lattices represented in Fig. S3 g-i, to adjust the results to the real experimental values.